\documentstyle[sprocl,epsfig]{article}

\def\Journal#1#2#3#4{{#1} {\bf #2}, #3 (#4)}

\def\NIMA{{\em Nucl. Instrum. Methods} A}

\def\PLB{{\em Phys. Lett.}  B}

\def\be{\begin{equation}}
\def\ee{\end{equation}}
\def\bea{\begin{eqnarray}}
\def\eea{\end{eqnarray}}

\def\GeV{\hbox{\rm{Ge\kern -0.1em V}}}
\def\MeV{\hbox{\rm{Me\kern -0.1em V}}}

\newcommand{\gev}   {\GeV}

\newcommand{\gevcc} {\GeV$\!/c^2$}

\newcommand{\sqs}   {\mbox{$\sqrt{s}$}}

\newcommand{\mh}    {\mbox{$m_{\mathrm h}$}}
\newcommand{\mH}    {\mbox{$m_{\mathrm H}$}}
\newcommand{\mHpm}  {\mbox{$m_{\mathrm H^\pm}$}}
\newcommand{\mA}    {\mbox{$m_{\mathrm A}$}}
\newcommand{\mW}    {\mbox{$m_{\mathrm W}$}}
\newcommand{\mZ}    {\mbox{$m_{\mathrm Z}$}}

\newcommand{\lplm}  {\mbox{$\ell^+\ell^-$}}
\newcommand{\epem}  {\mbox{$\mathrm  {e^+e^-}$}}
\newcommand{\mpmm}  {\mbox{$\mathrm  {\mu^+\mu^-}$}}
\newcommand{\tptm}  {\mbox{$\mathrm  {\tau^+\tau^-}$}}
\newcommand{\nunu}  {\mbox{$\mathrm  {\nu\overline{\nu}}$}}
\newcommand{\qqbar} {\mbox{$\mathrm  {q\overline{q}}$}}
\newcommand{\qqbarprime }{\mbox{$\mathrm  {q'\overline{q}'}$}}

\newcommand{\bbbar} {\mbox{$\mathrm  {b\overline{b}}$}}

\newcommand{\cscs}  {\mbox{$\mathrm  {c\bar{s}s\bar{c}}$}}
\newcommand{\tncs}  {\mbox{$\mathrm   {c\bar{s}}\tau^-\bar{\nu}_{\tau}$}}

\newcommand{\tntn}  {\mbox{$\tau^+\nu_{\tau}\tau^-\bar{\nu}_{\tau}$}}
\newcommand{\btn}   {\mbox{${\cal B}(\mathrm {H^+}\to\,\tau^+\nu_\tau)$}}

\newcommand{\ie}    {\hbox{\it i.e.}}

\newcommand{\eg}    {\hbox{\it e.g.}}
\newcommand{\cf}    {\hbox{\it cf.}}

\renewcommand{\to}   {\mbox{$\rightarrow$}}
\newcommand{\ipb}   {\mbox{$\mathrm {pb^{-1}}$}}

\begin{document}

\title{HIGGS SEARCHES AT LEP2 WITH THE ALEPH DETECTOR}

\author{E. KNERINGER}

\address{Institut f\"ur Experimentalphysik, Universit\"at Innsbruck,
6020 Innsbruck, AUSTRIA\\
ALEPH Collaboration\\
E-mail: Emmerich.Kneringer@uibk.ac.at} 


\maketitle\abstracts{
Data collected with the ALEPH detector at centre-of-mass energies
between 130 and 189 GeV are used to search for Higgs bosons
of the standard model and its supersymmetric extensions.
No evidence for a Higgs particle has been found in 256 \ipb\ of LEP2 data.
Mass exclusion limits were set.
}

\section{Introduction}\label{sec:intro}

\begin{figure}[ht]
\begin{tabular}{cc}
\epsfig{figure=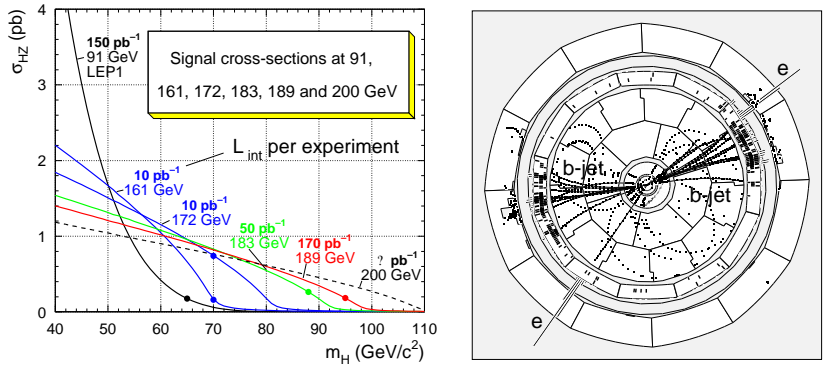,height=5.0cm} &
\epsfig{figure=dc044577_009277_zz.eps,height=4.9cm}
\end{tabular}
\caption{\label{fig:smh}
Left: cross section for the Higgs-strahlung process
at the various LEP energies.
For a given energy the point on the curve indicates
the standard model Higgs boson mass exclusion limit obtained
by a typical LEP experiment with the data taken at that energy.
Right: Higgs boson candidate event in the H$\ell\ell$ channel
(as seen by the ALEPH detector).
}
\end{figure}

A neutral Higgs boson is required to complete the particle spectrum
of the standard model. Fits to electroweak data~\cite{EWWG99}
from LEP, SLC and Tevatron indicate that
if the Higgs particle exists it has most probably a mass around 100 \gevcc.
The situation at LEP, which is summarized in Fig.~\ref{fig:smh} (left),
shows that we are getting more and more sensitive to this mass range.

All mass exclusion limits given in this report
are at 95\% confidence level.
To get an idea of the sensitivity of a specific analysis
the expected mass exclusion limit~\cite{CL95} from gedanken experiments
usually is also quoted.
All the search analyses use cut-based selections
and/or neural network methods.
We make extensive use of Monte Carlo simulations
for the neural net training and the setting of the selection cuts.
No details of the preliminary analyses are given,
however, the references always point to the corresponding
latest published analysis.

\section{The Standard Model Higgs Boson}\label{sec:sm}

At LEP2, the dominant Higgs production mechanism is the so-called
Higgs-strahlung process \epem \to\ Z$^*$\to\ HZ\@.
There are also minor contributions from processes where the Higgs
boson is generated through the fusion of two electroweak gauge bosons.
In the mass range accessible at LEP2, the Higgs boson decays in
92\% of the cases into two quark jets (91\% of which are \bbbar)
and in about 8\% of the cases into \tptm.
Combining these numbers with the branching fractions
of the Z boson we obtain the branching fractions of HZ into the
following four characteristic event topologies:
\begin{itemize}
\item four jet channel: \qqbarprime \qqbar\ ~(64.6\%)
\item missing energy channel: H\nunu\ ~(20.0\%)
\item lepton channel: H\lplm, $\ell$ = e or $\mu$ ~(6.7\%)
\item $\tau$--channel: either H or Z decays to \tptm\ ~(8.7\%)
\end{itemize}
Except for the lepton channel, a high b-quark content is required
for the quark jets coming from Higgs boson decay.
In all channels the invariant mass of the decay products
of the Z must be compatible with \mZ.
The results of the HZ analyses~\cite{SM184} applied to data taken
with the ALEPH detector~\cite{ALEPH90}$^,\,$\cite{ALEPH95} 
at 189 \gev\ are summarized in Table~\ref{tab:sm}.

\begin{table}[ht]
\caption{\label{tab:sm}
Signal efficiencies $\epsilon$, expected number of signal events $N^{exp}_{95}$
for \mH\ = 95~\gevcc, background expectation from standard model processes
$N^{exp}_{bkg}$, background from ZZ events $N^{exp}_{ZZ}$
(dominant background source; irreducible)
and number of observed events in 170 \ipb\ of ALEPH data
taken at \sqs\ = 189 \gev\ for the HZ decay topologies.
}
\vspace{0.2cm}
\begin{center}
\footnotesize
\begin{tabular}{|c|ccccc|}
\hline
\raisebox{0pt}[9pt][5pt]{Selection}  &
$\epsilon$ & $N^{exp}_{95}$ & $N^{exp}_{bkg}$ & $N^{exp}_{ZZ}$ & $N_{obs}$ \\
\hline
\bbbar\qqbar   & 39.3\% & 7.7 & 18.5 & 10.1 & \raisebox{0pt}[9pt]{20} \\
\bbbar\nunu    & 24.3\% & 1.6 &  3.7 &  3.1 &  7 \\
\bbbar\lplm    & 72.9\% & 1.5 & 13.2 & 11.8 & 13 \\
\bbbar\tptm    & 20.1\% & 0.2 &  0.8 &  0.6 &  2 \\
\tptm\qqbar    & 18.9\% & 0.3 &  1.8 &  1.2 &  0 \\
\hline
total          & {}    & 11.3 & 38.0 & 26.8 & 42 \\
\hline
\end{tabular}
\end{center}
\end{table}

The analyses for the four individual channels are optimized
in such a  way that the combination of these analyses
gives the best global HZ searches analysis.
The definition used here for `best analysis' is:
the analysis which gives the highest expected Higgs mass exclusion limit
under the hypothesis that there is no signal~\cite{CL95}.

\begin{figure}[ht]
\begin{tabular}{cc}
\epsfig{figure=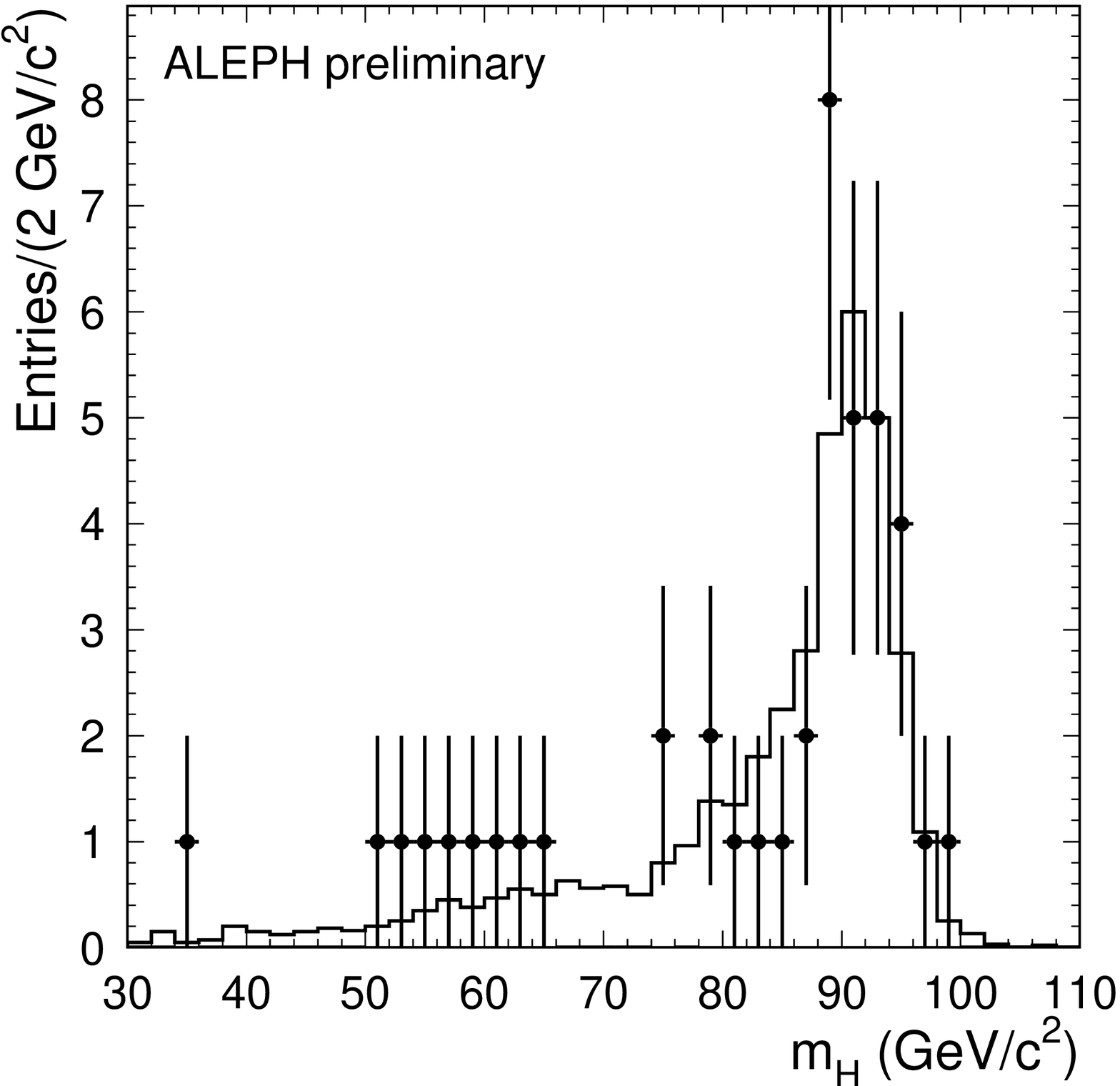,height=5.5cm} &
\epsfig{figure=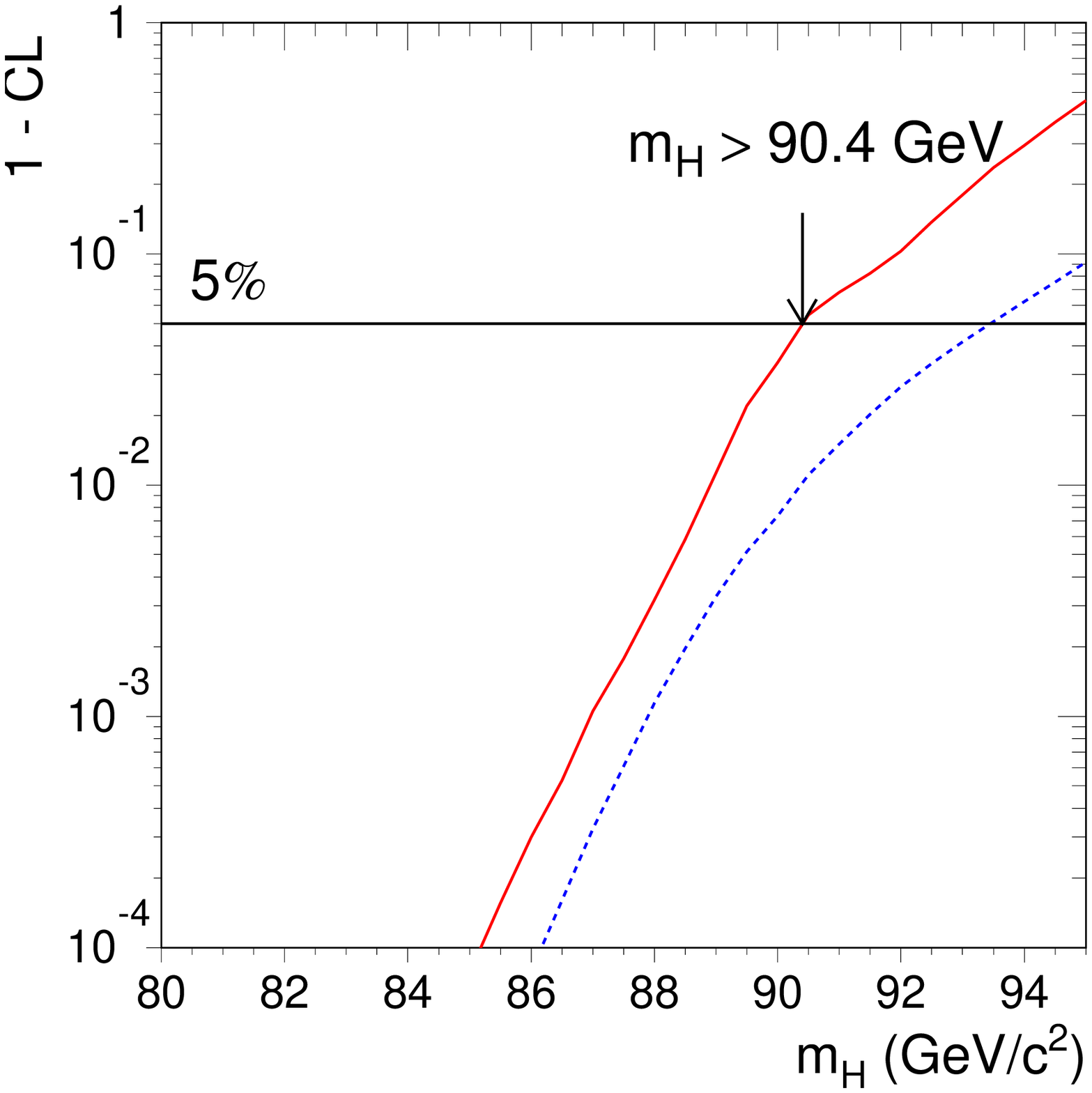,height=5.6cm}
\end{tabular}
\caption{\label {fig:sm}
Left: distribution of \mH\ of the selected events (points)
and the expectation from the standard model background processes (histogram)
in HZ searches at \sqs\ = 189 \gev.
Right: observed (full line) and expected (dashed line)
combined confidence levels for the standard model Higgs search.
}
\end{figure}

No signal is observed in 250 \ipb\ of data
(\cf\ \mh\ distribution in Fig.~\ref{fig:sm})
which allows to exclude a standard model Higgs boson
with a mass less than 90.4 \gevcc,
as shown in Fig.~\ref{fig:sm} (right).
The expected limit is 93.4 \gevcc.

\section{Neutral Higgs Bosons of the MSSM}\label{sec:hzha}

The particle spectrum of the Higgs sector of the minimal supersymmetric
extension of the standard model (MSSM)
consists of five physical states,
two CP-even neutral bosons h and H (with mixing angle $\alpha$
and masses \mbox{\mh\ $<$ \mH}), one CP-odd neutral boson A
and two charged bosons H$^\pm$.
At tree level, the Higgs sector
can be parameterized by two independent parameters,
\eg\ \mh\ and $\tan\beta$ = $v_2/v_1$, the ratio of the vacuum expectation
values of the two Higgs doublets.
Only the h and A bosons are within the reach of LEP2.
They are expected to be produced in Z decays via
the Higgs-strahlung process \mbox{Z$^*$\to\ hZ},
with a cross section proportional to $\sin^2(\beta-\alpha)$,
and via the associated pair production \mbox{Z$^*$\to\ hA},
with a cross section proportional to $\cos^2(\beta-\alpha)$.

\noindent
These two processes are complementary, in the sense
that if one cross section is maximal the other is minimal and vice versa.
Thus both processes must be searched for.
For the first process, the results from the hZ searches
(Sec.~\ref{sec:sm}) can be used.
The second process is searched for in the following two decay channels:
\begin{itemize}
\item hA \to\ \bbbar\bbbar\  ~(85\%)
\item hA \to\ \bbbar\tptm, \tptm\bbbar\  ~(15\%)
\end{itemize}
The performance of the two analyses~\cite{MSSM184} is reported
in Table~\ref{tab:hzha}.
\begin{table}[ht]
\caption{\label{tab:hzha}
Performance of the hA analyses at \sqs\ = 189 \gev:
signal efficiencies,
expected number of signal events for \mh\ = \mA\ = 85 \gevcc,
background expectation and number of observed events.
}
\vspace{0.2cm}
\begin{center}
\footnotesize
\begin{tabular}{|c|cccc|}
\hline
\raisebox{0pt}[9pt][5pt]{Selection}  &
$\epsilon$ & $N^{exp}_{85}$ & $N^{exp}_{bkg}$ & $N_{obs}$ \\
\hline
\bbbar\bbbar   & 58.3\% & 3.4 &  9.0 & \raisebox{0pt}[9pt]{13} \\
\bbbar\tptm    & 24.0\% & 0.2 &  0.6 &  0 \\
\hline
\end{tabular}
\end{center}
\end{table}
\begin{figure}[ht]
\begin{tabular}{cc}
\epsfig{figure=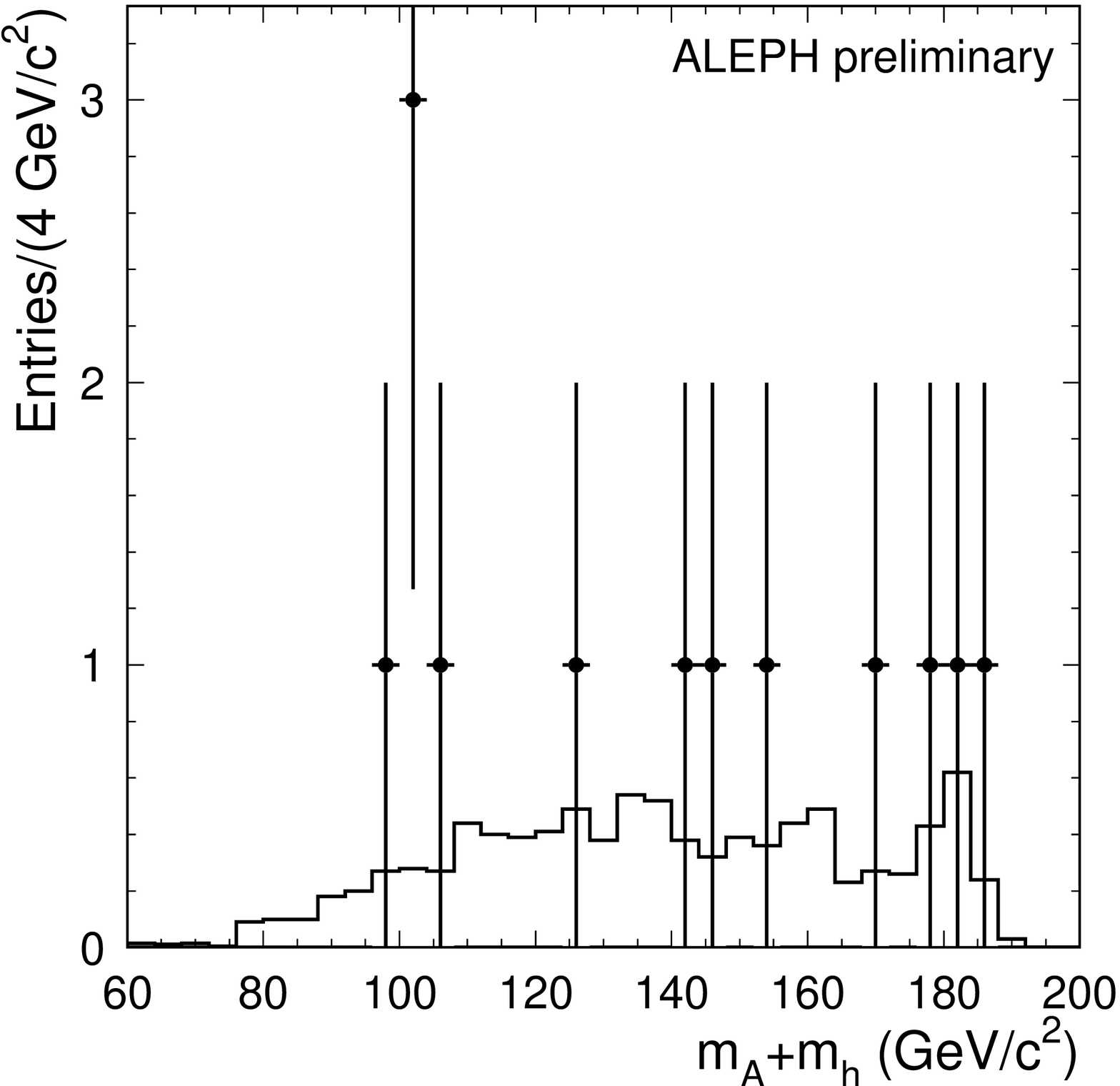,height=5.5cm} &
\epsfig{figure=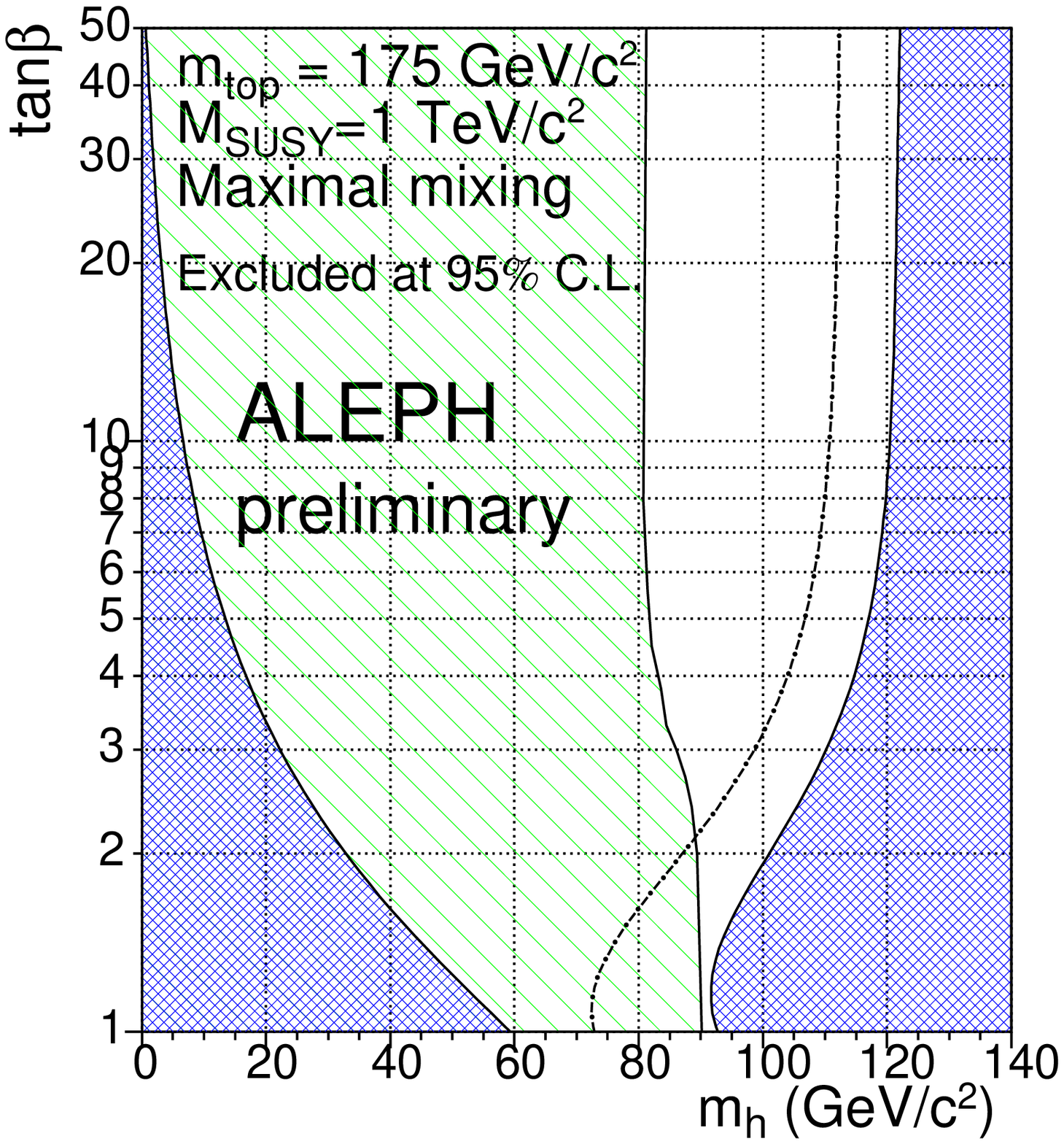,height=5.6cm}
\end{tabular}
\caption{\label{fig:hzha}
Left: distribution of \mA$+$\mh\ of the selected events (points)
and the expectation from the standard model background processes (histogram)
in hA searches at \mbox{\sqs\ = 189 \gev}.
Right: experimentally excluded region
in the $[$\mh,$\tan\beta ]$ plane
(after combination with the hZ searches).
The dark area represents
the theoretically forbidden region
for the case of `maximal stop mixing'
(dashed-dotted curve for the `no stop mixing' case).
}
\end{figure}

No signal is observed in the data,
as can be seen in Fig.~\ref{fig:hzha} (left).
This allows to set an upper limit on the cross section for hA production
as a function of \mh,
which implies an upper limit on $\cos^2(\beta-\alpha)$ as a function of \mh.
On the other hand, the hZ searches interpreted in this way result in an upper
limit on $\sin^2(\beta-\alpha)$ as a function of \mh.
Combining the hZ and hA searches in an optimal way~\cite{CL95}
leads to an excluded region
in the two-dimensional parameter space $[$\mh, $\sin^2(\beta-\alpha)]$,
which usually is translated to the $[$\mh, $\tan\beta]$ plane,
as shown in Fig.~\ref{fig:hzha} (right).
For $\tan\beta \ge 1$ we find that 
ALEPH data up to centre-of-mass energies of
189 \gev\ exclude the h and A Higgs bosons of the MSSM
with masses less than 80.8 and 81.2 \gevcc, respectively,
\ie\ the neutral Higgs bosons must be heavier than the W boson
if $\tan\beta \ge 1$.
\section{Charged Higgs Bosons}\label{sec:hphm}

In the MSSM charged Higgs bosons are heavier than W bosons.
This restriction does not hold for general two doublet models,
which are very attractive theoretically because of the absence
of flavor changing neutral currents and the relation
\mW\ = \mZ\ $\cos \theta_W$
holding at tree level.
The H$^\pm$ has the same decay modes as the W$^\pm$, but since its
coupling is proportional to the charged fermion masses,
it dominantly decays into the heaviest energetically allowed fermion pair
of the quark and lepton families.
Whether H$^\pm$ decay preferentially into quarks or leptons
depends on other parameters of the model.
Therefore, the search for pair-produced H$^+$H$^-$ is performed in 
the following three channels:
\begin{itemize}
\item leptonic channel: \tntn
\item mixed channel: \tncs 
\item hadronic or four jet channel: \cscs
\end{itemize}
The performance of these analyses~\cite{CHA184} is summarized
in Table~\ref{tab:hphm}.
\begin{table}[ht]
\caption{\label{tab:hphm}
Efficiencies $\epsilon$, 
number of standard model background events expected $N^{exp}_{bkg}$ 
and number of observed candidates $N_{obs}$
for the three charged Higgs boson analyses at a centre-of-mass energy
of 189~\gev, as functions of the charged Higgs mass. 
For the four jet channel, numbers are quoted 
within a $\pm 3$~\gevcc\ window around the assumed Higgs boson mass.
}
\vspace{0.2cm}
\begin{center}
\footnotesize
\begin{tabular}{|c|ccc|ccc|ccc|} 
\hline
\mHpm     & \multicolumn{3}{|c|}{\tntn} & 
\multicolumn{3}{|c|}{\tncs} & \multicolumn{3}{|c|}{\cscs} \\
\cline{2-10}
(\gevcc) &
 $\epsilon$ (\%) & $N^{exp}_{bkg}$ & $N_{obs}$ & 
 $\epsilon$ (\%) & $N^{exp}_{bkg}$ & $N_{obs}$ & 
 $\epsilon$ (\%) & $N^{exp}_{bkg}$ & \raisebox{0pt}[9pt][5pt]{$N_{obs}$} \\
\hline
50& 33.5 & 15.5 & 20 & 35.6 & 9.4 & 11 & 40.1 & 15.1 & 18\\
55& 35.2 & 15.5 & 20 & 37.2 & 9.4 & 11 & 38.4 & 22.8 & 23\\
60& 38.2 & 15.5 & 20 & 37.4 & 9.4 & 11 & 37.6 & 28.0 & 19\\
65& 35.2 & 15.5 & 20 & 34.8 & 9.4 & 11 & 37.2 & 30.6 & 28\\
70& 39.9 & 15.5 & 20 & 28.1 & 9.4 & 11 & 36.4 & 32.7 & 35\\
75& 40.8 & 15.5 & 20 & 19.1 & 9.4 & 11 & 34.6 & 55.1 & 40\\
\hline
\end{tabular}
\end{center}
\end{table}
\begin{figure}[ht]
\begin{tabular}{cc}
\epsfig{figure=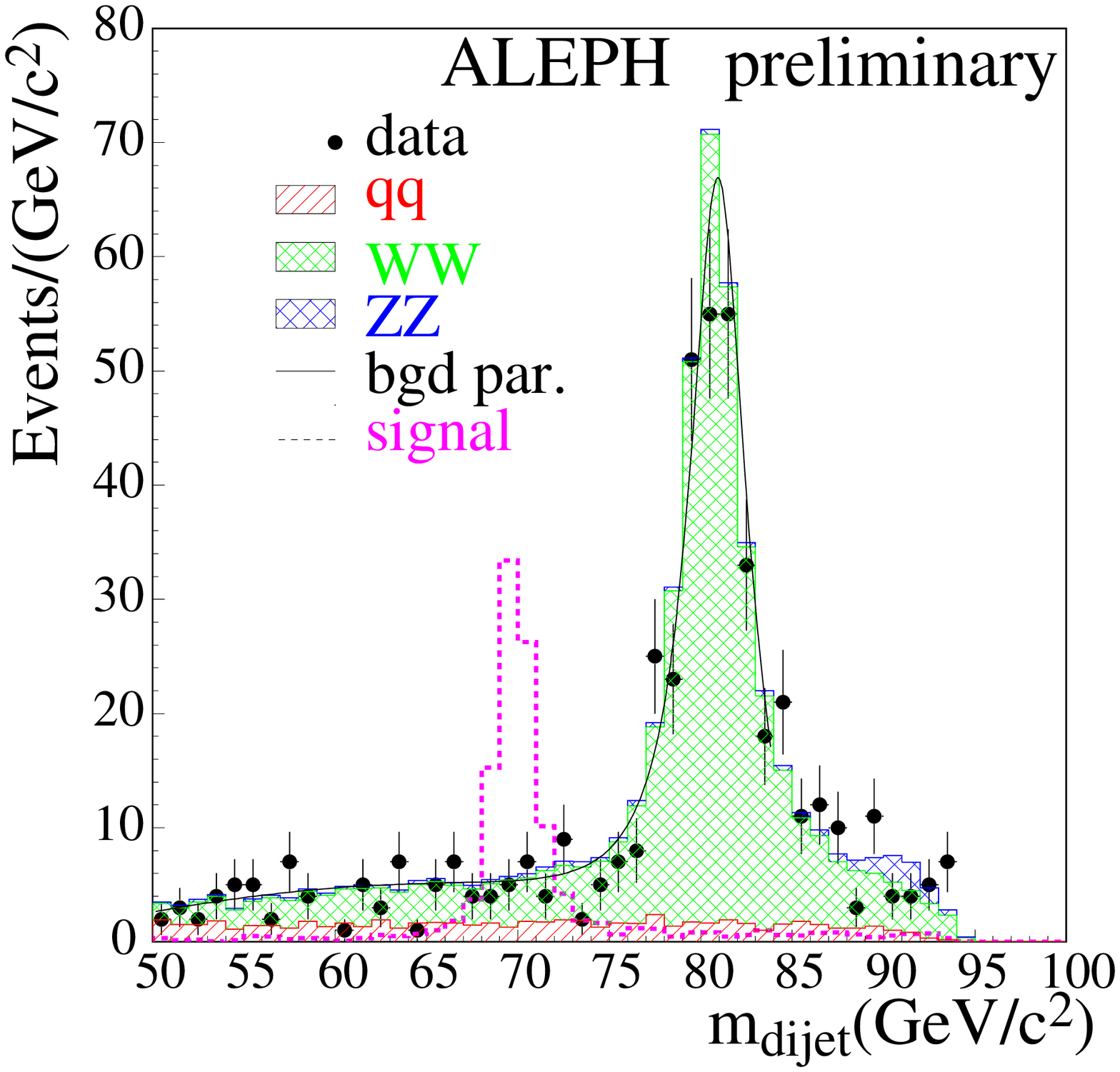,height=5.4cm} &
\epsfig{figure=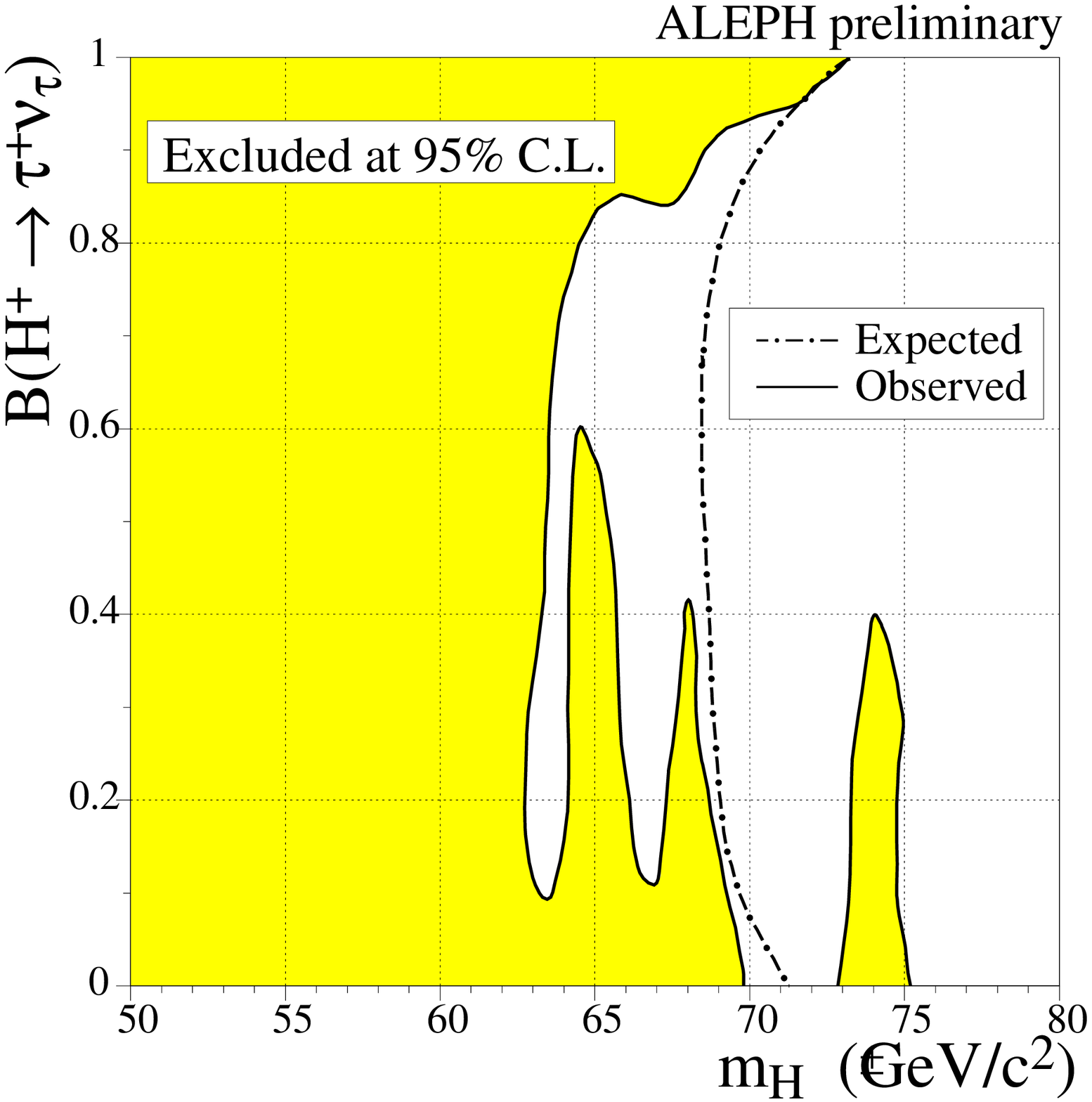,height=5.0cm}
\end{tabular}
\caption{\label{fig:hphm}
Left: distribution of the dijet mass distribution as obtained after
applying all cuts of the \cscs\ selection; shown are 189~\gev\ data
(dots), background Monte Carlo (histogram) and the
polynomial parameterization of the background (full line).
Also shown is the Monte Carlo expectation for a signal
with \mHpm = 70~\gevcc\ (dashed line) with arbitrary normalization.
Right: limit on the mass of charged Higgs bosons as a function
of \btn. Shown are the expected (dash-dotted)
and observed (full) exclusion curves for the combination of the three
charged Higgs boson analyses. The shaded area is excluded at 95\% C.L.
}
\end{figure}

The search for charged Higgs bosons in the three final states
\tntn, \tncs\ and \cscs\ has been performed using 175~\ipb\ of ALEPH data
collected at \sqs\ = 189~\gev. No evidence of Higgs boson production
was found (Fig.~\ref{fig:hphm} left) and mass limits were set as a function
of the branching ratio \btn.
The result of the combination of the three analyses is displayed in
Fig.~\ref{fig:hphm} (right) where the curves corresponding to expected and
observed confidence levels of 95\% exclusion are drawn.
As can be seen from this figure,
charged Higgs bosons with masses below 62.5~\gevcc\ are excluded at
95\%~C.L. independently of \btn,
where the expected mass exclusion limit is 68.5~\gevcc.

\section{Invisible Higgs Boson Decays}\label{sec:inv}
\begin{figure}[ht]
\begin{tabular}{cc}
\epsfig{figure=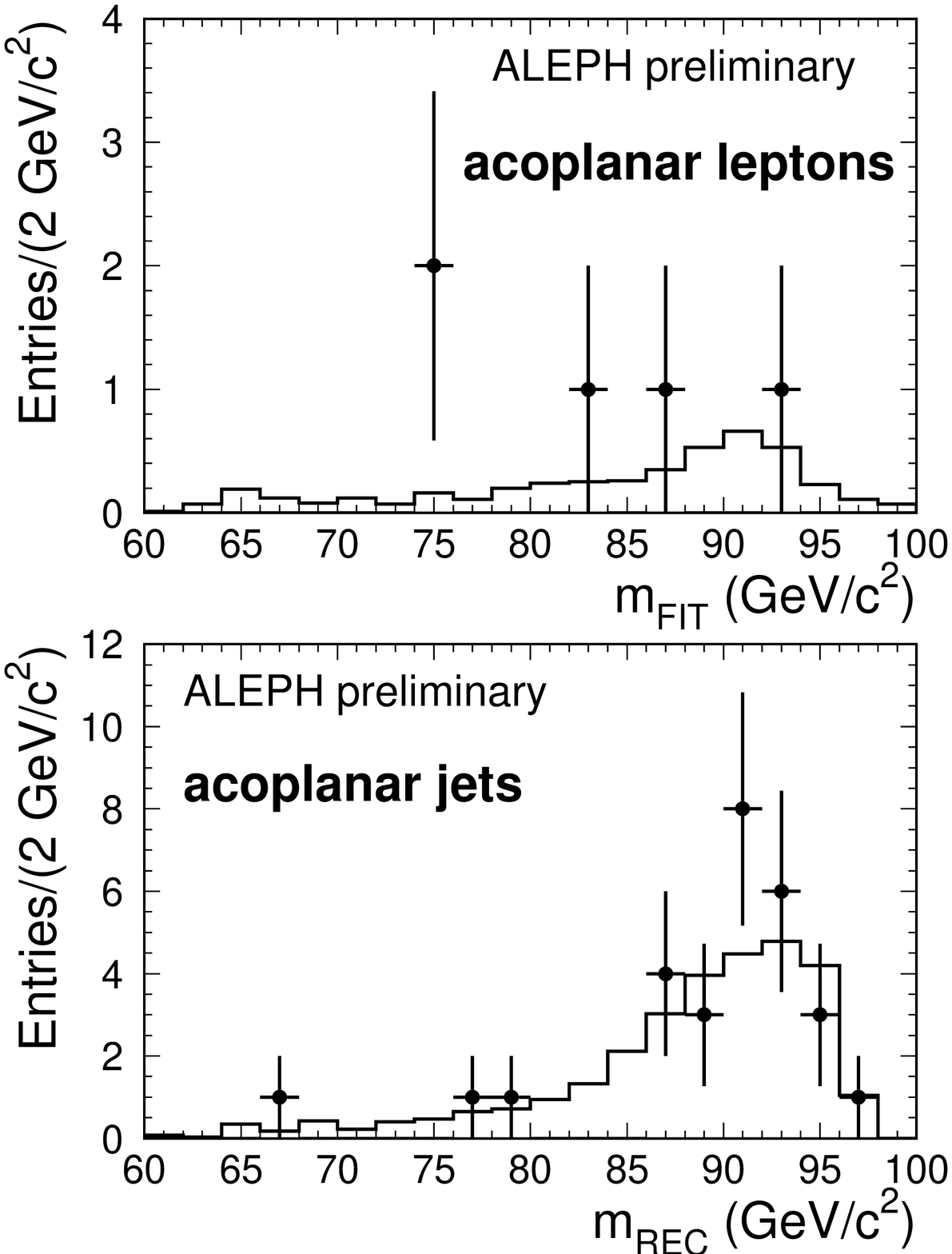,height=6.0cm} &
\epsfig{figure=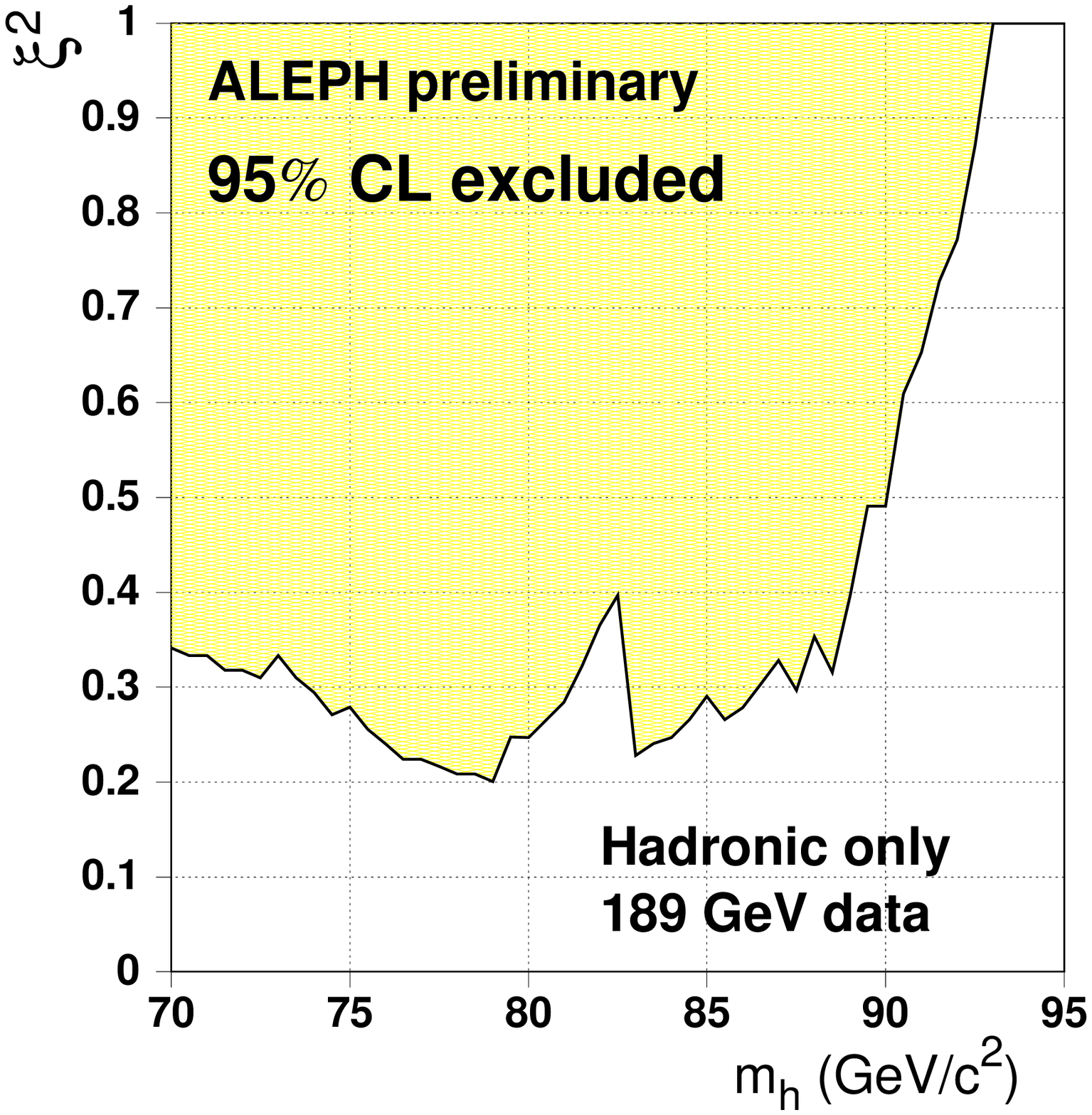,height=6.0cm}
\end{tabular}
\caption{\label{fig:inv}
Left: distribution of the reconstructed Higgs boson mass of the selected events
compared to the Monte Carlo expectation,
for the acoplanar lepton and jet pair searches.
Right: region in the $[$\mh, $\xi^2]$ plane excluded at 95\% C.L.
using only the hadronic channel at 189 \gev.
}
\end{figure}

Many extensions of the standard model allow for the Higgs boson
to decay invisibly, \eg\ into a pair of lightest neutralinos
when the neutralino $\chi$ is light enough.
Since the Higgs boson is produced through the Higgs-strahlung process hZ,
this leads to the following two event topologies:
\begin{itemize}
\item a pair of acoplanar leptons, when Z \to\ \epem\ or Z \to\ \mpmm\ 
\item a pair of acoplanar jets, when the Z decays hadronically
\end{itemize}
where the acoplanarity is defined as the azimuthal angle
between the two lepton or jet directions.
Two analyses~\cite{INVIS184} are designed for the two channels.
When they are applied to ALEPH data taken at \sqs\ = 189 \gev,
33 candidates are found, in agreement with 33.6 events expected
from all background processes (\cf\ Table~\ref{tab:inv}).
The distributions of the reconstructed Higgs boson masses are shown
in Fig.~\ref{fig:inv} (left).

\begin{table}[ht]
\caption{\label{tab:inv}
Performance of the acoplanar lepton and acoplanar jet pair analyses.}
\vspace{0.2cm}
\begin{center}
\footnotesize
\begin{tabular}{|c|cc|}
\hline
Selection  & $N^{exp}_{bkg}$ & \raisebox{0pt}[9pt][5pt]{$N_{obs}$} \\
\hline
h\lplm   &  4.3 & 5 \\
h\qqbar  & 29.3 & 28 \\
\hline
\end{tabular}
\end{center}
\end{table}
Quite generally,
the production cross section for invisible Higgs boson decay
can be parameterized as
$\xi^2 \cdot \sigma_{\mathrm {SM}}$(\epem \to\ hZ),
 where $\xi^2$ is a model dependent
factor ranging from 0 to 1.
One way of presenting the result of the negative searches is
to calculate the 95\% C.L. level upper limit on the production cross section
of the invisibly decaying Higgs boson
in units of $\sigma_{\mathrm {SM}}$(\epem \to\ hZ) as a function of \mh,
which is shown in Fig.~\ref{fig:inv} (right).

\section{Summary}

In this report, an overview has been given
of the present status of Higgs boson searches with the ALEPH detector.
The results are mainly based on the analysis of data taken at
189 \gev\ centre-of-mass energy.
When the data taken at lower energies are included in the analyses
the 95\% confidence level exclusion limits improve slightly.

For the standard model Higgs boson we find 
\mH\ $>$ 90.4 \gevcc\ at 95\% C.L.
using 250 \ipb\ of data taken at \sqs\ = 161, 172, 183 and 189 \gev.
This means that a Higgs boson with a mass equal to the Z mass just
cannot be excluded at 95\% C.L. with ALEPH data taken up to the year 1998.

Including all lower energy data, for the neutral
Higgs bosons of the MSSM we obtain the limits
\mh\ $>$ 80.8 \gevcc\ and \mA\ $>$ 81.2 \gevcc\ (valid for all values
of $\tan\beta \ge 1$).

From the analysis of 175 \ipb\ of ALEPH data collected in 1998,
the charged Higgs bosons of general two doublet models
are excluded below 62.5 \gevcc\ 
independent of their decay mode.

Finally, using the same data,
invisibly decaying Higgs bosons are searched for.
For a production cross section equal to that of the standard model Higgs boson,
\ie\ $\xi^2 = 1$,
masses below 92.8 \gevcc\ are excluded.

\section*{Acknowledgments}
I would like to thank all my colleagues from the ALEPH
Higgs Task Force for the continuous effort to
produce all these results.


\section*{References}
\begin{thebibliography}{99}
\bibitem{EWWG99}LEP Electroweak Working Group, CERN-EP/99-15.
\bibitem{CL95}P. Janot and F. Le Diberder, \Journal{\NIMA}{411}{449}{1998}.
\bibitem{SM184}ALEPH Collaboration, \Journal{\PLB}{440}{403}{1998}.
\bibitem{ALEPH90}ALEPH Collaboration, \Journal{\NIMA}{294}{121}{1990}.
\bibitem{ALEPH95}ALEPH Collaboration, \Journal{\NIMA}{360}{481}{1995}.
\bibitem{MSSM184}ALEPH Collaboration, \Journal{\PLB}{440}{419}{1998}.
\bibitem{CHA184}ALEPH Collaboration, \Journal{\PLB}{450}{467}{1999}.
\bibitem{INVIS184}ALEPH Collaboration, \Journal{\PLB}{450}{301}{1999}.

\end{thebibliography}
\end{document}